# A line code with quick-resynchronization capability and low latency for the optical data links of LHC experiments

**Binwei Deng**[a,b], **Mengxun He**[c], **Jinghong Chen**[d,e], **Di Guo**[f,b], **Suen Hou**[g], **Xiaoting Li**[h,b], **Chonghan Liu**[b], **Ping-Kun Teng**[g], **Annie C. Xiang**[b], **Yang You**[e], **Jingbo Ye**[b], **Datao Gong**[b*], **Tiankuan Liu**[b*]

[a] *Hubei Polytechnic University, Huangshi, Hubei 435003, P. R. China*
[b] *Department of Physics, Southern Methodist University, Dallas, TX 75275, USA*
[c] *University of Texas at Dallas, Richardson, TX 75080, USA*
[d] *Department of Electrical and Computer Engineering, University of Arizona, Tucson, AZ 85721, USA*
[e] *Department of Electrical Engineering, Southern Methodist University, Dallas, TX 75275, USA*
[f] *University of Science and Technology of China, Hefei, Anhui 230026, P. R. China*
[g] *Institute of Physics, Academia Sinica, Nangang 11529, Taipei, Taiwan*
[h] *Central China Normal University, Wuhan, Hubei 430079, P.R. China*
    E-mail: dtgong@mail.smu.edu, tliu@mail.smu.edu

ABSTRACT: We propose a line code that has fast resynchronization capability and low latency. Both the encoder and decoder have been implemented in FPGAs. The encoder has also been implemented in an ASIC. The latency of the whole optical link (not including the optical fiber) is estimated to be less than 73.9 ns. In the case of radiation-induced link synchronization loss, the decoder can recover the synchronization in 25 ns. The line code will be used in the ATLAS liquid argon calorimeter Phase-I trigger upgrade and can also be potentially used in other LHC experiments.

# Contents

## 1. Introduction

Optical links have been extensively used in Large Hadron Collider (LHC) experiments [1-2] because of their advantages of high bandwidth, high channel density, low mass, and no ground loop. A typical application of an optical link in LHC experiments is shown in Figure 1. On the transmitter side, an optical transmitter converts the electrical signal to the optical signal, which is transmitted through an optical fiber from the detector to the counting room. A serializer is used to multiplex parallel data into serial data and transmit through a single optical fiber. On the receiver side, an optical receiver recovers the electrical signal from the fiber. A deserializer recovers the parallel data from the serial data. In an optical link, data from multiple channels are multiplexed and transmitted through a single optical fiber. A line code must be used to process the data before transmission. The line code keeps the serial data to be DC balanced and limits the length of consecutive identical digits (CIDs) in order for the receiver clock-data recovery (CDR) circuit to recover the clock. The line code also provides boundary identification and error detection or correction. The line code is implemented on the transmitter side in an encoder shown in figure 1. Correspondingly, a decoder on the receiver side recovers the original user data and check if any error happens during the data transmission. The encoder and the serializer are usually integrated in a single component called the transmitter, while the decoder and the deserializer are integrated in a single component called the receiver.

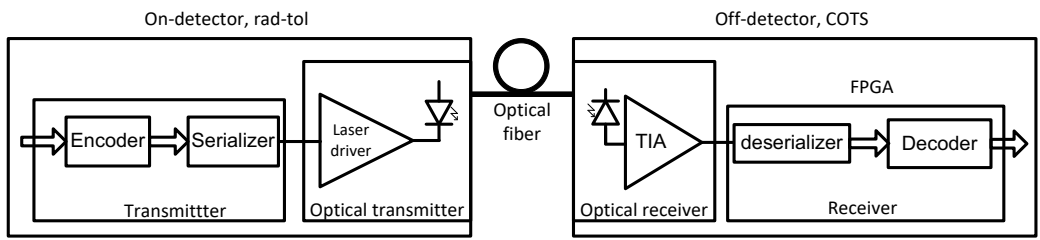

Figure 1: A block diagram of a typical optical link in LHC experiments.

When an optical link is used in LHC experiments, it must meet extra requirements. In the LHC experiments, all components on the transmitter side of the data link are mounted on the detector and must tolerate harsh radiation environment [3]. Therefore, most components on the transmitter side, including the serializer and the encoder, are custom designed to be radiation-tolerant. All components on the receiver side are located in the counting room and are not exposed to radiation. The receiver circuits can thus be implemented with Commercial-Off-The-Shelf (COTS) components. For a line code used in LHC experiments, the encoder is typically implemented in a radiation-tolerant Application-Specific Integrated Circuit (ASIC), whereas the decoder can be implemented in a commercial Field-Programmable Gate Array (FPGA). We have designed and tested two serializer ASIC prototypes [4-5] in a commercial 0.25-μm Silicon-on-Sapphire (SoS) CMOS technology for the ATLAS Liquid Argon Calorimeter Phase-I trigger upgrade [6]. The SoS CMOS technology is chosen because it is immune to the single event latchup and has smaller single event upset cross section than equivalent bulk technologies. During the single event effect test of a serializer ASIC prototype, synchronization losses were observed in additional to single bit upsets. A synchronization loss began with a burst of bit errors in a short duration of less than 80 unit intervals (UIs) and followed by the received data stream consistently shifted one bit earlier or later [4]. In order to minimize the data loss, the fast resynchronization capability is highly desirable for the line code used in LHC experiments. Moreover, the encoder of the line coder as well as the corresponding serializer must be suitable to be implemented in a radiation-tolerant ASIC.

In addition to the challenging radiation tolerance requirement, latency is an important requirement when an optical link is used in the trigger system. In general, a shorter latency is preferred so that the event buffer used to store the data can be smaller. When a few sub-detector readout systems are upgraded whereas other sub-detectors remain unchanged, the latency of the new sub-detectors must be no more than that of the existing sub-detectors. This is the case of the ATLAS Liquid Argon Calorimeter Phase-I trigger upgrade [6]. The latency budget of the optical link designed for the ATLAS LAr calorimeter trigger upgrade is 150 ns, not including the time passing through the optical fiber. The encoder and the decoder are the major contributors of the latency of the optical link.

In order to achieve the required quick resynchronization and low latency, the line code must be carefully chosen. After studying several line codes commonly used in industry, we found that none meets the requirements. For example, from the view point of ASIC design, the 8B/10B line code [7] is difficult to be implemented. The 8B/10B encoder contains large lookup tables, which consume a significant amount of power and area. The serializer matching the 8B/10B encoder is a 10:1 serializer that consists of a 5:1 multiplexer and a 2:1 multiplexer. In the divider chain of the Phase-Locked Loop (PLL), a divided-by-5 circuit is needed to provide clock signals for the 5:1 multiplexer. At very high data rates, implementing the 5:1 multiplexer and the divided-by-5 circuit in the 0.25-μm SoS CMOS technology is challenging. Another example is the 64B/66B line code. Per the IEEE 10G-Ethernet standard, it takes a long time (64 × 66 = 4224 serial bits) to achieve resynchronization once the synchronization is lost.

We propose a custom line code, dubbed as LOCic, for the optical data links of the ATLAS liquid argon calorimeter trigger upgrade and other LHC experiments with latency and radiation tolerance requirements. The line code has the features of low latency and quick resynchronization capability. Both the encoder and decoder have been verified in FPGAs. The encoder has also been implemented in an ASIC. The performances of the line code are simulated in the case of ASIC implementation and measured in the case of FPGA implementation.

The remainder of the paper is organized as follows: Section II describes the proposed line code definition. The designs of the encoder and the decoder are discussed in Sections III and IV, respectively. Section V discusses the latency performance based on the ASIC simulation and the FPGA measurements. Section V summarizes the paper.

## 2. The line code definition

We assume the input data of LOCic encoder come from 8 ADC channels as a typical case in high-energy physics experiment, but LOCic is not limited to ADC data. All ADC channels sample and digitize the analog signals with a 40-MHz LHC bunch crossing clock at the resolution of 12 bits. The input data format of the ASIC ADCs [8] is shown in Figure 2. The bits $D_0$-$D_{11}$ in each channel are the digitized data of each channel. The order of the digitized data can be either Most Significant Bit (MSB) first or Least Significant Bit (LSB) first. The bits $D_{12}$-$D_{13}$ in each channel carry calibration data. The bits $T_0$-$T_{15}$ of all 8 channels are dummy and not used in the ADC. The 12-bit digitized data, the 2-bit calibration data, and the 2-bit dummy data of each channel are serialized and then sent out from each ADC channel at the data rate of 640 Mbps. For each ADC chip, a data clock of 320 MHz is provided to capture the data conveniently. In the COTS ADC implementation [9], no calibration data are required. The output data rate of the ADC is 480 Mbps with the data clock being 240 MHz. Two extra dummy bits in each channel are inserted to every digitized data of the 12 bits and the data rate is increased from 480 Mbps to 560 Mbps. In both ASIC and COTS implementations, a frame clock with the same frequency (40 MHz) of the LHC bunch crossing clock is provided to identify the beginning of each sample.

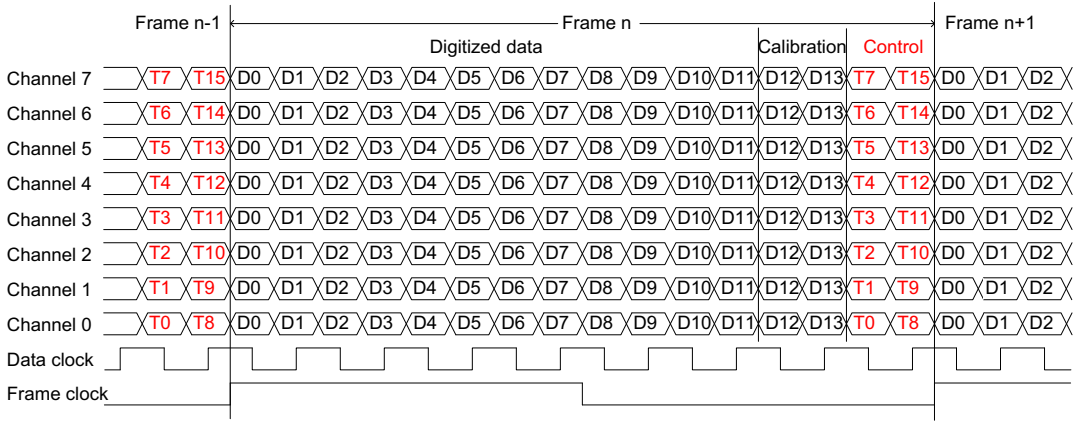

Figure 2: The LOCic data frame definition.

We define that the data input in one LHC bunch crossing clock cycle as a frame. Each frame consists of the digitized data, the calibration bits, and the extra dummy bits. In the LOCic encoder, the extra dummy bits in each frame are replaced by the frame control code. Each frame control code consists of 16 bits and is separated into CRC, frame boundary and PRBS fields.

The control code $T_0$−$T_7$ is an 8-bit cyclic redundant checking (CRC) code for the receiver to check if all data are transmitted correctly. The 8-bit CRC code is calculated from the raw data before they are scrambled. We choose the polynomial $P(x) = x^8+ x^5+ x^3+ x^2+ x^1+ x^0$ to calculate the CRC code [10]. Based on the selection of the scrambling polynomial which will be described later, it is proved that a single-bit error which occurs in the serializer after the scrambler generates three single-bit errors in the descrambled data sequence. The CRC polynomial we chose has the maximum Hamming distance with the CRC length of 8 bits and the message size of 112 bits and is capable of detecting three single-bit errors in a frame. The synchronization loss with up to 80 bits of burst errors is beyond the capability of the CRC protection.

The control code $T_8-T_{11}$ is the frame boundary field which is always "1010." Since the frame clock will not be transmitted through the optical fiber, there must be a mechanism for the receiver to identify the boundary of each ADC sample and the boundary of each ADC channel. The field $T_8-T_{11}$ serves as the frame boundary. The field also limits the length of CIDs to be no more than the frame length. Limited CID length is required for an optical link in which AC-coupling is used.

The remainder four bits ($T_{12}T_{13}T_{14}T_{15}$) is the PRBS field that provides the Bunch Crossing Identification (BCID) for the receiver to identify one frame from another and align different channels. The PRBS field also imposes extra restriction on the frame boundary. The bits $T_{12}T_{13}T_{14}T_{15}$ are separated into two pairs ($T_{12}T_{13}$ and $T_{14}T_{15}$). The first pair ($T_{12}T_{13}$) are taken from a $2^5-1$ PRBS beginning with "11 00 01 10 11 10 10 10 …," two bits per frame. The pair $T_{12}T_{13}$ is "11" in Frame 0 (i.e. BCID = 0), "00" in Frame 1 (BCID = 1), "01" in Frame 2 (BCID = 2), and so on. The formation of $T_{12}T_{13}$ is shown in Figure 3. The period of the $2^5-1$ PRBS is 31 bits. By using two bits in each frame, the pair $T_{12}T_{13}$ will repeat after 31 frames. The pair $T_{14}T_{15}$ is taken from a $2^7-1$ PRBS beginning with "11 00 00 00 10 …", two bits per frame. The period of the $2^7-1$ PRBS is 127 bits. Also by using two bits as each frame, the pair $T_{14}T_{15}$ will repeat after 127 frames. It can be proved that the four bits $T_{12}T_{13}T_{14}T_{15}$ repeats after 3937 (=31×127) frames. The period is larger than 3564, the period of the BCID. After BCID reaches its period, the field $T_{12}T_{13}T_{14}T_{15}$ will be reset to its initial value at Frame 0 and the period of the field $T_{12}T_{13}T_{14}T_{15}$ is trimmed to the period of the BCID.

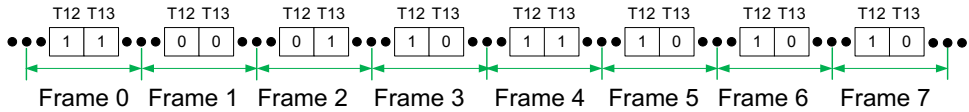

Figure 3: The formation of the field $T_{12}T_{13}$.

The control code $T_8-T_{15}$ is defined as the end of a frame. In practice, once the position of the control code $T_8-T_{15}$ is found, the start of the next fame will be known. Therefore, we can also consider the control code $T_8-T_{15}$ as the start of a frame.

The serial data going through the optical fibers need to be DC-balanced. The raw data will be scrambled before transmitted whereas the frame control code will not be scrambled. We choose the scrambling algorithm used in the 10-Gigabit Ethernet standard, $G(x) = x^{59} + x^{39} + 1$. One disadvantage of scrambling is that specific input sequences can generate undesirable output sequences. However, the probability of having such input sequences is very low for the scrambler with a 58-degree polynomial. One special exception, which can be very common in the input sequence, is the all-zero input sequence. It can be proved that as far as the initial values (usually called "seeds") of the scrambler are not all zeroes, the scrambler can still function properly when the data in the input sequence are all zeroes. The non-zero seeds of the scrambler are stored in internal configuration registers which need to be set after each power cycle.

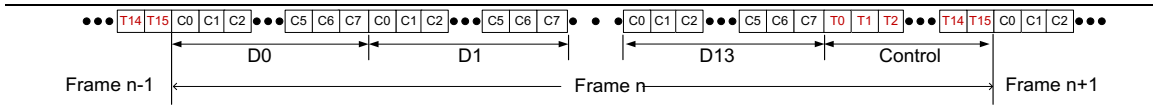

Figure 4: The serial data stream transmitted through the optical fiber. Channel 0, Channel 1, …, Channel 7 are shown as C0, C1, C2, …, C7, respectively.

After the dummy bits in each frame are replaced by the control code and the raw data are scrambled, the data are sent to the serializer, which generates a serial data stream. The serial data stream going through the optical fiber is shown in Figure 4. A protocol in which Channel 0 is the first transmitted channel and Channel 7 is the last channel to be transmitted is adopted in the serializer.

Each frame has 128 bits in the case of the ASIC ADC and 112 bits in the case of the COTS ADCs, respectively. For the frame with calibration data, the payload is 112 bits and the frame control field is 16 bits, so the overhead is 14.3%. For the frame without calibration data, the payload is 96 bits and the frame control field is 16 bits, and therefore, the overhead is 16.7%.

## 3. The encoder implementation

The encoder of the proposed line code has been implemented in an ASIC prototype fabricated in the 0.25-μm SOS CMOS technology and two FPGAs: Altera Stratix II GX and Xilinx Kintex-7.

The block diagram of the ASIC encoder is shown in Figure 5. The encoder consists of a synchronous First-In-First-Out (FIFO) buffer, a PRBS generator, a CRC generator, a scrambler, and a frame builder. The synchronous FIFO buffer is designed to use a COTS ADC to test the ASIC ADCs that are not available at the moment. When the data come from an 8-channel COTS ADC, each sample has 12 bits at 40 MHz and the data rate is 480 Mbps. The FIFO buffer adds 4 extra dummy bits for every 12 bits and changes the data rate to 640 Mbps. After the FIFO buffer, the data are processed synchronously with an internal 640-MHz clock. The PRBS generator is used to generate the frame control code $T_8$-$T_{15}$. The CRC generator generates the CRC data. The framer builder combines the framer control code and the scrambled data.

To minimize the latency, the encoder maintains the format of serial data of ADCs and adds extra frame control code in each frame to build 8-bit-wide data frame for the following 8:1 serializer. This design eliminates data buffers to orientate the data, reducing the latency and the hardware cost. In ASIC implementation, the design is optimized to operate as fast as possible in order to reduce the latency. The pipeline technique is used to achieve the speed as high as 640-MHz.

One advantage of the SoS CMOS technology is that the single event upset (SEU) cross section is lower than the equivalent bulky CMOS technologies. Based on the SEU test results of the previous serializer prototype, we have not used any special design techniques in the encoder ASIC to mitigate radiation induced SEU errors. The SEUs, which are supposed to be rare, in internal counters or state machines are flushed out in the next data frame and will not generate any single event functional interrupt. The internal configuration registers, which stores a non-zero seed of the scrambler and are missing in the current encoder prototype, will be protected with Triple Modular Redundancy (TMR) in the final design. The SEU cross section of the prototype encoder ASIC will be evaluated and special techniques such as TMR will be applied in the final design if necessary.

The ASIC has been submitted for fabrication. The ASIC occupies an area of 1.2 mm × 1.0 mm. Post-layout simulations show that the power consumption of the ASIC is about 200 mW. In the future, the encoder will be integrated in an ASIC together with an 8:1 serializer [5].

The FPGA implementation of the encoder is similar to the ASIC implementation except that the operation clock is 320 MHz in FPGAs. The encoder implemented in FPGAs will not be used in the real applications, but serves two purposes: (1) the verification of the function blocks of the ASIC encoder; (2) the data source of the decoder before the ASIC encoder is available.

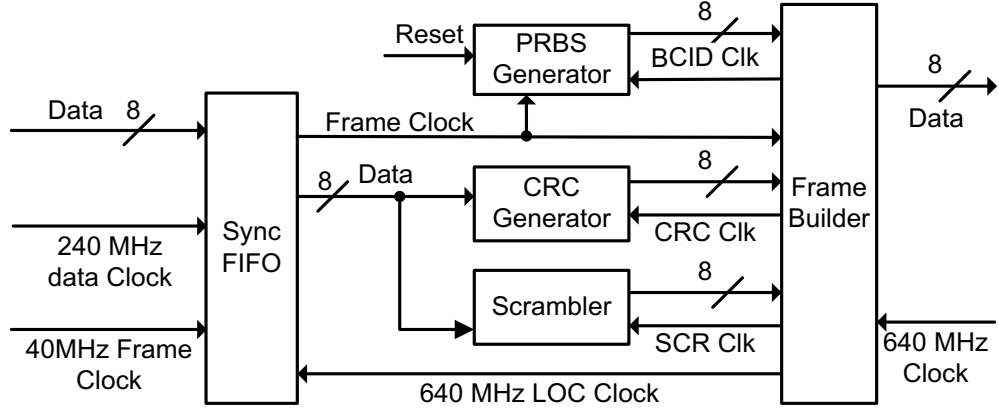

Figure 5: the block diagram of the encoder ASIC.

## 4. The decoder implementation

The proposed line code decoder has been implemented in two FPGAs: Altera Stratix II GX and Xilinx Kintex-7. The block diagram of the firmware on the receiver side in Kintex-7 is shown in Figure 6. All function blocks operate at a 320-MHz clock coming out of the deserializer, which has an input reference clock of 320 MHz. The deserializer converts the serial data stream into 16-bit parallel data. The synchronizer searches for the frame boundary through a state machine. The data extractor extracts the frame data to different fields based on the frame boundary identified by the synchronizer. The BCID generator calculates the 12-bit BCID from the PRBS field in the frame control code. The descrambler recovers the original raw data. The CRC checker verifies the CRC field of the recovered raw data in each frame. The output data of the decoder include the recovered data, a 12-bit BCID, a CRC flag indicating if the received data are error free, and a frame flag indicating whether the data are valid. The latency on the receiver side is critical and depends on two components, the deserializer and the decoder. For the deserializer, we reduce the latency on the receiver side by bypassing all unnecessary function blocks. For the decoder, we raise the operation clock frequency as high as possible to reduce the latency. In Stratix II, the operating clock is 160 MHz at the data width of 32 bits, resulting in a larger latency than in Kintex-7. Therefore, a Kintex-7 FPGA is preferred in the view point of latency.

The synchronization process that identifies the received frame boundary is implemented in the synchronizer and its state machine is shown in Figure 7. The synchronization process is described as follows. In the Check state, the state machine searches for the frame boundary fields and the PRBS fields in the data stream at an initial frame boundary pointer. If "1010" is found at the identical position of 4 consecutive frames and the PRBS fields of these 4 consecutive frames comply with the PRBS generation rule, the synchronization is achieved and the state machine goes to the Sync state. Otherwise, the state machine stays in the Check state for next search in which the frame boundary pointer shifts one bit. In the Sync state, the state machine keeps comparing the boundary and PRBS fields of the incoming frame with the predicted frame control code. If the received control code matches the predicted code, the state stays in the Sync state. Otherwise, the link synchronization may get lost due to a single event upset and the state machine goes into the Re-Sync state. In the Re-Sync state, the state will go back to the Sync state in either of the following two conditions:

(1) The frame boundary field and the PRBS field of the next frame matches the predicted values;
(2) With one bit of shift, either left or right, the frame boundary field and the PRBS field of the next frame matches the predicted values.

These two conditions are the synchronization loss scenarios found in the single event upset test [4]. After the serializer recovers from radiation-induced synchronization loss, the descrambler can produce valid data in 59 bits and the re-synchronization, which does not rely on the descrambler because the frame control field is not scrambled, can be achieved in a single frame (i.e., 128 bits or 25 ns) by simply keeping original frame boundary position or with one-bit shift. If none of the above two conditions is met, the state machine goes to the Check state to start over the synchronization process.

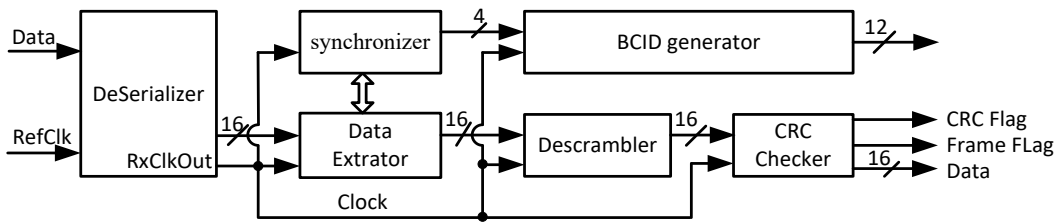

Figure 6: The block diagram of the FPGA implementation on the receiver side

The BCID is calculated as soon as the synchronizer reaches the SYNC state. The 4-bit PRBS field in a single frame cannot independently determine a 12-bit BCID. However, the PRBS fields in the consecutive four frames can be combined to form two bytes, one from the PRBS5 field and the other from the PRBS7 field. Once these two PRBS bytes are formed, the PRBS fields in all the following frames can be uniquely determined. Therefore, the relative position of a PRBS field in the series of consecutive PRBS fields can be determined by these two PRBS bytes. We use the relative position of a PRBS field in the series of consecutive PRBS fields to represent the BCID. The mapping from the two PRBS bytes to the BCID is complicated and it is difficult to be implemented in a single look-up table. To overcome this difficulty, we simplify the design by implementing only a mapping subset which haves the following 8 BCIDs: 0, 496, 992, 1489, 1984, 2480, 2976, and 3472. The PRBS5 bytes of these 8 BCIDs have the same value. The mapping to these 8 BCIDs is implemented in a lookup table. The BCID is initially set to an invalid value when the decoder is not in the SYNC state or the BCID calculation is not completed. If the PRBS5 byte or the PRBS7 byte does not fall in the subset and the BCID of the previous frame is valid, the BCID will increase by one. If the BCID of the previous frame is invalid, the BCID remains invalid. The drawback of this simplified implementation is that in the worst case, it takes the time equivalent to 495 frames to calculate the first valid BCID after the first Sync state is achieved from the Check state. However, this long search process occurs only once when the Sync state is achieved from the Check state.

When the Sync state is achieved from the Re-Sync state, it takes only the time equivalent to one frame (i.e. 25 ns) to calculate the BCID from the previous valid BCID.

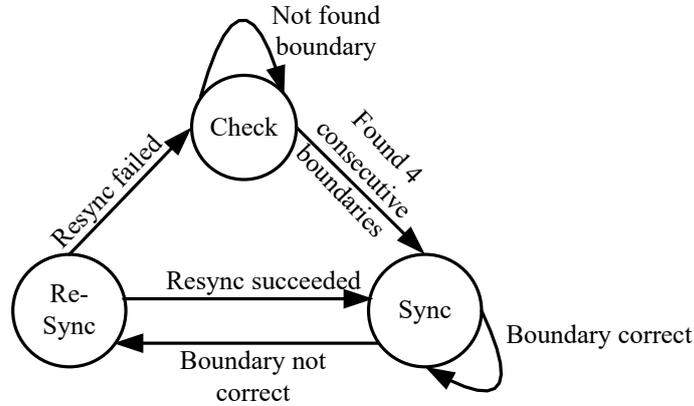

Figure 7: The state machine of frame synchronization process

In the Kintex-7 implementation, the encoder uses 395 registers, 384 lookup tables, and 135 slices, and the decoder uses 365 registers, 522 LUTs, and 183 slices.

## 5. The latency performance

The latency of the encoder implemented in an ASIC has been simulated. The latencies of each functional block of the encoder and the decoder implemented in FPGAs can be conveniently measured by using the ChipScope Pro Analyzer tool. The latencies of the serializer and the whole link are measured using a high-speed real-time oscilloscope. We calculate the latency of the deserializer by subtracting the latency of the encoder, the decoder, and the serializer from the latency of the whole link.

Table 1: The latency of the optical link

| Function block | | Latency (ns) | |
| --- | --- | --- | --- |
| | | Kintex-7 | ASIC + Kintex-7 |
| transmitter | FIFO | 3.1-6.3 | 1.6-3.1 |
| | Scrambler & CRC generator | 3.1 | 1.6 |
| | Frame Builder | 3.1 | 1.6 |
| | Serializer | 14.4 | 4.7 |
| | Total | 23.8-26.9 | 9.4-10.9 |
| receiver | Deserializer | 28.5-31.4 | |
| | Data Extractor | 9.4 | |
| | Descrambler | 3.1 | |
| | CRC Check | 3.1 | |
| | Total | 44.1-47.0 | |
| Total | | 67.9-73.9 | 53.5-57.9 |

The latency of the whole link with the transmitter implemented in an ASIC and the receiver implemented in an FPGA is estimated based on the latency measurements of Kintex-7 FPGA and the latency simulation of the transmitter ASIC. The latency of the optical link is summarized in Table 1. The latency of the whole link with the transmitter implemented in an ASIC and the receiver implemented in Kintex-7 is no more than 57.9 ns. The latency of the whole link with both the transmitter and the receiver implemented in Kintex-7 is no more than 73.9 ns. In both cases, the latency is less than half of the requirement, leaving enough design margins.

The latency shown in Table 1 is not fixed for the FIFO and the deserializer. The latency variation comes from the phase uncertainty of the internal clocks (the 640-MHz clock shown in Figure 5 and the 320-MHz clock shown in Figure 6) after each power cycle. The total latency variation is small enough to be absorbed when the data are latched with the 40-MHz LHC bunch crossing clock before the data are sent to the trigger system. The latency of the whole link is guaranteed to be fixed at the scale of the 40-MHz LHC bunch crossing clock.

## 6. Conclusion

We propose a line code that has fast resynchronization capability and low latency. Both the encoder and the decoder have been verified in FPGAs. The encoder has been designed in an ASIC prototype and will be integrated in an ASIC with an 8:1 serializer in the future. The latency of the whole optical link (not including the optical fiber) is estimated to be less than 73.9 ns. In the case of radiation-induced link synchronization loss, the decoder can recover the synchronization in 25 ns. The line code will be used in the ATLAS liquid argon calorimeter Phase-I trigger upgrade and can also be potentially used in other LHC experiments with similar requirements.

## Acknowledgments

This work is supported by US-ATLAS R&D program for the upgrade of the LHC, the US Department of Energy Grant DE-FG02-04ER1299, and the National Science Council in Taiwan. We are grateful to Drs. Jaroslav Ban, Gustaaf Brooijmans, Bill F Sippach of Nevis Laboratories of Columbia University, Drs. Daniel Dzahini and Benjamin Trocme of LPSC Grenoble, Drs. Hucheng Chen, Kai Chen and Hao Xu of Brookhaven National Laboratory, Dr. Stefan Simion of LAL Orsay Laboratory, Drs. Nicolas Dumont Dayot and Guy Perrot of LAPP, Drs. Jinhong Wang and Junjie Zhu of University of Michigan at Ann Arbor, Drs. Ken Johns and Bill Hart of University of Arizona for beneficial discussions.